\documentclass[epsf,prb,twocolumn,showpacs]{revtex4-1}
\usepackage[pdftex]{graphicx}
\usepackage{dcolumn}
\usepackage{bm}
\usepackage{epsfig}
\usepackage{latexsym}
\usepackage{amsmath}
\usepackage{amsfonts}
\usepackage{amssymb}
\usepackage{color}
\usepackage{array}
\usepackage{framed}
\usepackage{esvect}
\usepackage{caption}
\usepackage{subcaption}

\renewcommand{\(}{\left(}
\renewcommand{\)}{\right)}

\newcommand{\be}{\begin{equation}}
\newcommand{\ee}{\end{equation}}
\newcommand{\bea}{\begin{eqnarray}}
\newcommand{\eea}{\end{eqnarray}}

\newcommand{\p}{\partial}

\newcommand{\la}{\langle}
\newcommand{\ra}{\rangle}

\newcommand{\lp}{\left(}
\newcommand{\rp}{\right)}

\def\nn{\nonumber\\}

\begin{document}

\title{Bound states of three fermions forming symmetry-protected topological phases}
\author{Chong Wang}
\affiliation{Department of Physics, Massachusetts Institute of Technology,
Cambridge, MA 02139, USA}
\date{\today}
\begin{abstract}
  We propose a simple theoretical construction of certain short-range entangled phases of interacting fermions, by putting the bound states of three fermions (which we refer to as {\em{clustons}}) into topological bands. We give examples in two and three dimensions, and show that they are distinct from any free fermion state. We further argue that these states can be viewed as combinations of certain free fermion topological states and bosonic symmetry-protected topological (SPT) states. This provides a conceptually simple understanding of various SPT phases, and the possibility of realizing them in cold atom systems. 
New parton constructions of these SPT phases in purely bosonic systems are proposed.
 We also discuss a related anomaly in two dimensional Dirac theories, which is the gravitational analogue of the parity anomaly.

\end{abstract}
\maketitle



Topological insulators (TIs) are gapped phases of matter hosting non-trivial boundary states protected by symmetries. Most of our current knowledge of topological insulators come from free fermion models\cite{TIs}, which have been fully classified\cite{tenfold} in all dimensions and with different global symmetries.

Recently an interesting generalization of the free fermion topological insulators to interacting systems - known as Symmetry-protected topological (SPT) phases - has been pursued theoretically (see Ref. \onlinecite{review} for simple review articles). These are states with a gap and no fractionalization in the bulk, but nevertheless have interesting surface states protected by global symmetries. In contrast to more exotic phases such as the fractional quantum hall states, SPT phases are {\em short-range entangled} and do not posses intrinsic topological order\cite{Wenbook}. In systems of free fermions, the notion of SPT states simply reduces to that of topological band insulators.

It was realized\cite{chencoho2011} that SPT states can also exist in systems of interacting bosons. A classic example is the Haldane spin-$1$ chain, which is not fractionalized in the bulk and hosts degenerate end states protected by symmetries. The boson SPT states can also be realized in interacting fermion systems, since one can always bind two fermions to form a boson, such as the electron spin $\frac{1}{2}c^{\dagger}\sigma c$ or the Cooper pair $c_{\uparrow}c_{\downarrow}$ - a process which clearly requires fermionic interactions. One can then imagine putting the bound bosons into a boson SPT state.

It should be noted that this approach does not always give new states. Some of the boson SPT states become equivalent to certain free fermion states\cite{note0} once physical fermions are introduced into the system. The corresponding free fermion states could be trivial\cite{trivialhaldane,TScSTO,3dfSPT} or topological\cite{luashvin,TScSTO}. A simple example is that with only time-reversal symmetry, the Haldane chain becomes equivalent to four copies of the Kitaev chain\cite{1dfSPT} with spinless fermions.

There are, however, many other boson SPT states that are distinct from free fermion models. Abundant examples have been found in both 2D\cite{luashvin} and 3D\cite{3dfSPT, 3dfSPT2}. So far, these states have been exclusively understood within the bosonic approach: in all the models the non-triviality comes entirely from the boson sector (spins, Cooper pairs, etc.), and the existence of fermions does not seem to contribute anything.

In this work, we show that some of these non-trivial boson SPT states can also be understood in an intrinsically fermionic approach, even though they are distinct from any free fermion state. Specifically, these states can be viewed as topological insulators of certain fermions - not the free ones, but the bound states of three fermions (or some other odd number) which we refer to as {\em{clustons}}. This observation not only provides new insights into the interacting SPT states, but also suggests realizations in cold atom systems: three-body bound state can be achieved in cold atom systems through Efimov effect\cite{efimov}, thus if one can control these Effimov states efficiently and put them in a topological band (which is certainly quite challenging), it will be possible to realize the novel states we propose. 

We study specifically two examples in this paper. The first one is the boson integer quantum hall state (BIQHE) in 2D\cite{tsml}, the second one is the boson topological insulator (BTI) in 3D\cite{avts12,hmodl,metlitski}. In both examples the charged bosons are viewed as Cooper pairs of electrons. One can also view the fermions as slave-particles (partons), and our result gives another way to write down wavefunctions of these states in purely bosonic systems.

\section{2D: clustons in Chern band}
\label{2D}

We consider a fermion system with charge $U(1)$ symmetry. Now imagine a situation in which fermions prefer to form three-body bound states (clustons), which clearly requires strong interaction. We then put the clustered fermions into a Chern band with Chern number $C=1$. This state is not fractionalized, so one could naturally ask if it is equivalent to a free fermion state. We answer this question by looking at its transport properties.

The quantum hall conductance is given by $\sigma_{xy}=\tilde{e}^2C=9$ in units of $e^2/h$, where $\tilde{e}=3$ is the charge carried by the clustered fermion. However, the thermal hall conductance $\kappa_{xy}$ in units of $\frac{\pi^2}{3}\frac{k_B^2}{h}T$, also known as the chiral central charge, is given by $\kappa_{xy}=C=1$, since the thermal transport is independent with the amount of charge carried by the fermions. Therefore this state is distinct from all the integer quantum hall states made of free fermions, which always have $\sigma_{xy}=\kappa_{xy}$ since heat carriers are also charge carriers in a free fermion systems. 
For interacting systems made of electrons (fermions with charge $e$), it is known\cite{luashvin} that as long as the bulk excitation spectrum does not contain fractionalized anyons (i.e. the excitations only include fermions with odd-integer charge and bosons with even-integer charge), the difference between the hall conductance and the thermal hall conductance is always an integer multiple of eight in proper units:
\be
\label{diff}
\frac{3}{\pi^2}\frac{h}{k_B^2T}\kappa_{xy}-\frac{h}{e^2}\sigma_{xy}=8n.
\ee
For completeness we reproduce the derivation of this result in Appendix \ref{0}. Our cluston Chern insulator is thus a minimal state with unequal charge and thermal hall conductance.

\begin{figure}[ttt]
\includegraphics[width=1.2in]{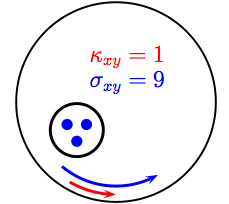}
\caption{Construction of the cluston Chern insulator: putting the three-fermion clustons into a Chern band.}
\label{2dcluster}
\end{figure}

To make the structure clearer, we combine the above state with a free fermion IQHE with Chern number $\bar{C}=-1$. The total system now has $\sigma_{xy}=8$ but $\kappa_{xy}=0$. 
We now consider another state with the same transport properties: imagine the fermions form Cooper pairs (charge $e^*=2$ bosons), and the Cooper pairs form a boson IQHE state. It was shown in Ref. \onlinecite{luashvin,tsml,liuwen} that such a state would be non-chiral ($\kappa_{xy}=0$), but with quantum hall conductance $\sigma_{xy}=2(e^*)^2=8$. The transport properties of the Cooper pair BIQHE state therefore matches perfectly with the state we constructed above. In Appendix \ref{a} we use the edge theory to show explicitly that the two states are indeed equivalent, even though they appear to be very different in the way they are constructed. More explicitly, the Cooper pair BIQHE state can be described by the simple fermionic Hamiltonian:
\be
\label{13chern}
H[f,F]=H_{C=1}[F]+H_{C=-1}[f]+\sum_{ijkl}\lambda F_i^{\dagger}f_jf_kf_l+h.c.,
\ee
where $f$ denotes the charge-$1$ fermion, $F$ denotes the charge-$3$ clustered fermion, $i,j,k,l$ represent indices such as spins and sub-lattices, $H_C$ is a quadratic Hamiltonian that puts the fermions into a band with Chern number $C$, and the last term reveals $F$ as the bound state of three fundamental fermions.

One can also consider a different state, where $F$ is in a band with $C=1$, while $f$ is in a band with $\bar{C}=-9$. The total Hall conductance is then $\sigma_{xy}=C(e^*)^2+\bar{C}e^2=3^2-9=0$, but the thermal hall conductance is $\kappa_{xy}=C+\bar{C}=-8$. The transport signature is identical to that of the $E_8$ state\cite{luashvin}, which is the minimal chiral state of charge-neutral bosons. In systems of charge-neutral fermions, the $E_8$ state can be understood as $16$ copies of $p+ip$ superconductors\cite{kitaev}. Our work provides another way to understand the state in terms of charged fermions.

\section{3D: clustons in topological band}
\label{3D}

We now consider a three dimensional fermion system with $U(1)$ charge conservation and time-reversal symmetry $\mathcal{T}$ with $\mathcal{T}^2=-1$ on the physical fermions. Again imagine a situation in which fermions prefer to form three-body bound states (clustons). We then put the charge-$3$ clustons into a Fu-Kane-Mele topological band\cite{FKM}. The state is obviously interacting and not fractionalized, so one can again ask what phase the state belongs to. It is easy to see that the state should be nontrivial, for example, through the magneto-electric response\cite{qi} described by a $\theta$-term with 
\be
\theta=\pi(e^*)^2=9\pi=\pi\hspace{5pt}({\rm mod} 2\pi).
\ee
A naive answer could then be that the state is equivalent (connected) to the free fermion topological insulator. However, we will show that this is not true.

\begin{figure}
\begin{subfigure}{.2\textwidth}
\centering
\includegraphics[width=.7\linewidth]{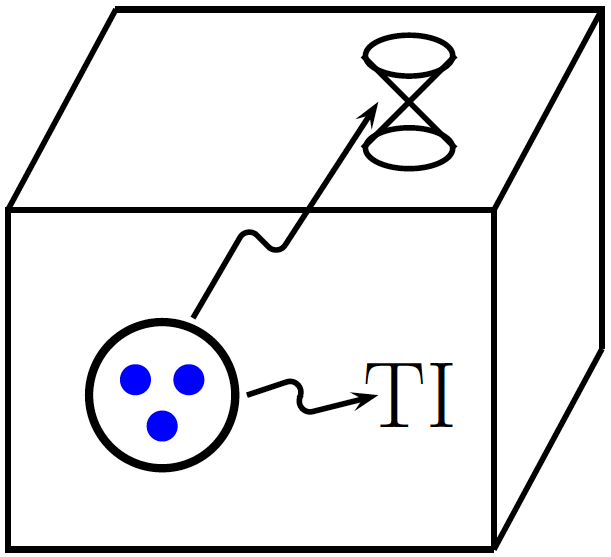}
\caption{}
\label{cluster}
\end{subfigure}%
\begin{subfigure}{.2\textwidth}
\centering
\includegraphics[width=.9\linewidth]{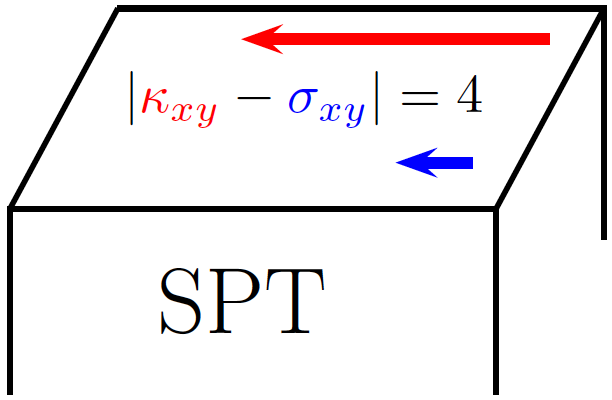}
\caption{}
\label{transport}
\end{subfigure}
\caption{(a) Construction of the cluston TI: putting the three-fermion clustons into a topological band. (b) Transport signature on a $\mathcal{T}$-breaking surface. $\sigma_{xy}-\kappa_{xy}=4({\rm mod} 8)$ signifies a nontrivial SPT.}

\end{figure}

We answer the question by looking at the surface. The simplest symmetric surface state is a single Dirac cone of fermions carrying charge $e^*=3$. However, for our purpose it is easier to reveal the nontriviality (and fully determine the topological state) from symmetry-breaking surface states. If we break the $U(1)$ charge conservation on the surface (e.g. by depositing a superconductor on top), the Dirac cone can be gapped out. 
If we keep the $U(1)$ symmetry but break time-reversal $\mathcal{T}$ instead, we can also gap out the Dirac cone, but the surface will have nontrivial transport signatures. It is well known that the Dirac mass gap leads to ``half'' quantum hall and thermal hall conductance. In our case we have $\sigma_{xy}=\frac{1}{2}(e^*)^2=\frac{9}{2}$, and $\kappa_{xy}=\frac{1}{2}$ since the carrier charge does not affect thermal transport. We therefore have $\sigma_{xy}-\kappa_{xy}=4$, which is half of what is allowed in strictly two dimensions without fractionalization (for example, the state considered in Eq.~\eqref{13chern} has $\sigma_{xy}-\kappa_{xy}=\pm 8$).

The $\sigma_{xy}-\kappa_{xy}$ mismatch is zero on the surface of the free fermion topological insulator, since heat carriers are also charge carriers in free fermion systems. When interactions are introduced, the mismatch can shift by integer multiples of eight, by depositing either the BIQHE state of Cooper pairs, or the $E_8$ state of spins, {\em i.e.} $\sigma_{xy}-\kappa_{xy}=8n$ if the state is equivalent to the free fermion TI. Therefore the state we described above cannot be equivalent to the free fermion TI.

Non-fractional insulators (SPT states) with $U(1)$ and $\mathcal{T}$ symmetries in three dimensions are classified\cite{3dfSPT} by $\mathbb{Z}_2^3$. There are six nontrivial SPT states distinct from the free fermion TI. Three out the six states have $\sigma_{xy}=\frac{1}{2}({\rm mod}1)$ when $\mathcal{T}$ is broken on the surface. Among the three states, only one of them can be completely gapped out by breaking $U(1)$ charge-conservation (while keeping $\mathcal{T}$). This state also has $\sigma_{xy}-\kappa_{xy}=4({\rm mod} 8)$ when $\mathcal{T}$ is broken on the surface. Thus the clustered TI state described above is precisely this state. Ref. \onlinecite{3dfSPT} provided two equivalent ways to understand the state: one can think of it either as the combination of the free fermion TI and a spin SPT (dubbed topological paramagnet $e_fTm_fT$ in Ref. \onlinecite{hmodl,3dfSPT}), or as the combination of the free fermion TI and the bosonic topological insulator (BTI)\cite{avts12,hmodl,metlitski} of Cooper pairs. 
Our clustered TI provides another way to view the state, and it differs from the free fermion TI by a bosonic SPT state of either the spin or the Cooper pair.

Our result also provide a simple way to (at least theoretically) construct the BTI state with fermions, in the same spirit as Eq.~\eqref{13chern}:
\be
\label{13TI}
H[f,F]=H_{FKM}[F]+H_{FKM}[f]+\sum_{ijkl}\lambda F_i^{\dagger}f_jf_kf_l+h.c.,
\ee
where $f$ denotes the charge-$1$ fermion, $F$ denotes the charge-$3$ clustered fermion, $i,j,k,l$ represent indices such as spins and sub-lattices, $H_{FKM}$ is a quadratic Hamiltonian that puts the fermion into the Fu-Kane-Mele band, and the last term reveals $F$ as the bound state of three fundamental fermions. 

Previously, the simplest symmetry-preserving surface state of the topological paramagnet $e_fTm_fT$ was given by a gapped topologically ordered $\mathbb{Z}_2$ gauge theory $\{1,e,m,\epsilon\}$, with both the electric-like particle $e$ and magnetic-like particle $m$ being fermions and Kramers' ($\mathcal{T}^2=-1$). Likewise, the simplest symmetry-preserving surface state of the boson TI (BTI) was given by another gapped topologically ordered $\mathbb{Z}_2$ gauge theory $\{1,e,m,\epsilon\}$, with both the $e$ and $m$ particles carrying half-charge (in our case the Cooper pair boson carries charge-$2$, so $e$ and $m$ carry charge-$1$). The two states are distinct as purely bosonic states, but in the presence of electrons (charge-$1$ fermions), the two states become equivalent since the two surface gauge theories can be transformed to each other by attaching an electron to the $e$ and $m$ particles. 
Eq.~\eqref{13TI} leads to another simple symmetric surface state without topological order (but is gapless instead), namely two Dirac cones carrying charge $e=1$ and $e^*=3$ respectively:
\be
\label{2dirac}
\mathcal{L}=\bar{\psi}\sigma^{\mu}(-i\partial_{\mu}+A_{\mu})\psi+\bar{\Psi}\sigma^{\mu}(-i\partial_{\mu}+3A_{\mu})\Psi,
\ee 
where $A_{\mu}$ is the external probe gauge field. This also implies that the Dirac theory in Eq.~\eqref{2dirac}, even though is free from the famous parity anomaly\cite{paranomaly}, suffers from another anomaly first proposed by Vishwanath and Senthil\cite{avts12} in the context of topological quantum field theories. The Vishwanath-Senthil anomaly can be viewed as a gravitational analogue of the parity anomaly: if the theory in Eq.~\eqref{2dirac} is coupled to gravity, then a gravitational Chern-Simons term at level $c=4({\rm mod}8)$ must be introduced to regularize the theory\cite{note}, thus time-reversal symmetry must be broken. In terms of the bulk theory, this corresponds to a gravitational $\theta$-term\cite{graanomaly} at $\theta=8\pi$.

One can also show the equivalence between the fermionic state in Eq.~\eqref{13TI} and the Cooper pair BTI directly from the symmetric surface states. The idea is to show that the surface Dirac theory in Eq.~\eqref{2dirac} can be gapped by introducing the corresponding $\mathbb{Z}_2$ topological order. However, the argument, which we briefly outline in Appendix \ref{b}, is considerably more technical. The fact that the equivalence was easily established using the result in Ref. \onlinecite{3dfSPT} is another illustration of the usefulness of the $\mathbb{Z}_2^3$ classification.

One can also imagine similar states with five-fermion clustons, or any other odd number for $e^*$. By repeating the previous argument, it is easy to show that the cluston TI differs from the free fermion TI by the Cooper pair BTI if $e^*=\pm3 ({\rm mod} \hspace{2pt}8)$, and equivalent to the free fermion TI if $e^*=\pm1 ({\rm mod} \hspace{2pt}8)$.

\section{Bosonic states: parton constructions}

So far we have discussed various states in fermionic systems (i.e. systems with fermions in the microscopic Hilbert spaces). In this somewhat more technical section we consider purely bosonic systems (without fermions in the microscopic Hilbert space) by gauging the fermions. This leads us to some new parton constructions of various bosonic states.

\textbf{2D}: For the $E_8$ state, we start from nine copies of free fermion Chern insulator and one copy of charge-$3$ cluston Chern insulator with the opposite chirality. We then gauge the $U(1)$ symmetry. Since the state has no net hall conductance, the dynamics of the compact $U(1)$ gauge theory does not contain Chern-Simons term. It is well known that in $2+1$ dimensions a compact $U(1)$ gauge theory without Chern-Simons term is always confined. Therefore we automatically obtain a confined (unfractionalized) bosonic state with $\kappa_{xy}=8$, which has a chiral edge state with chiral central charge $c_+-c_-=8$. 

For the BIQHE state, we start from Eq.~\eqref{13chern} and then gauge the fermion parity ($f\to(-1)f$) which is a $\mathbb{Z}_2$ symmetry (constructions using higher gauge symmetries were proposed earlier in Ref. \onlinecite{2dparton}). A simple way to realize this in a parton construction is to start with two flavors of charge-$2$ bosons $B_{1,2}$ and decompose them as
\begin{eqnarray}
\label{parton}
B_1&=&f_1f_2, \nonumber \\
B_2&=&f^{\dagger}_3F,
\end{eqnarray}
where $f_{1,2,3}$ are charge-$1$ fermions and $F$ is a charge-$3$ fermion. It is possible to arrange a mean field ansatz for the fermions such that the gauge symmetry reduces to a simple $\mathbb{Z}_2$ which is the fermion parity, and $F$ form a band with Chern number $C=1$ while $f_{1,2,3}$ together form a band with Chern number $C^*=-1$. Since the state is non-chiral, the $\mathbb{Z}_2$ gauge flux (also dubbed as ``vison'') carries bosonic statistics. The total quantum hall conductance is $\sigma_{xy}=8$, so the vison carries integer charge and can always be neutralized by binding certain number of $f$ fermions (which does not change the vison statistics due to the mutual statistics). The system can thus go through a confinement transition by condensing the gauge flux, and the resulting state is a confined bosonic state (with bosons carrying charge $e^*=2$), with $\sigma_{xy}=8=2(e^*)^2$ and $\kappa_{xy}=0$. 

\textbf{3D}: We start from Eq.~\eqref{13TI}. If we gauge the $U(1)$ symmetry, the $U(1)$ gauge theory does not contain nontrivial $\theta$-angle ($\theta=9\pi-\pi=0({\rm mod} \hspace{2pt}2\pi)$), therefore the gauge theory can be confined while preserving $\mathcal{T}$. The resulting system has only charge-neutral bosons (spins), and we obtain the topological paramagnet dubbed $e_fTm_fT$ in Ref. \onlinecite{hmodl,3dfSPT}.

Instead of gauging the $U(1)$ symmetry, we can also choose to gauge the fermion parity ($f\to(-1)f$) which is a $\mathbb{Z}_2$ symmetry (for example using the parton decomposition in Eq.~\eqref{parton}). The $\mathbb{Z}_2$ gauge theory can again be confined, and we get a system of charge-$2$ bosons. The resulting state is then the BTI of these charge-$2$ bosons. Constructions with higher gauge symmetries were proposed earlier in Ref. \onlinecite{3dparton}. 

There are, however, two subtle issues on this construction. The first issue is that whether the $\mathbb{Z}_2$ flux loops coupled to the fermions in Eq.~\eqref{13TI} can indeed proliferate and produce a confined gapped bulk. This is nontrivial because naively the loop hosts gapless fermion modes\cite{z2defect}. We show in Appendix \ref{c} that the flux core can indeed be gapped, hence the flux loops can proliferate and confine the fermions. Since the gapless mode in a $\mathbb{Z}_2$ flux core in a $3D$ TI is identical to the edge mode of the $2D$ TI\cite{kanemele}, our result also shows that putting charge-$3$ clustons into a $2D$ TI (quantum spin hall state) does not produce any new state, instead it gives the conventional $2D$ TI. Interestingly, this is related to the absence of Cooper-pair boson SPT state in $2D$ fermion systems.

The second issue is the nature of the confined (bosonic) state. The boson system after confinement has $U(1)\rtimes\mathcal{T}$ symmetry. In such systems the $e_fTm_fT$ topological paramagnet and the bosonic topological insulator (BTI) are two distinct states, unlike in fermion systems where the two become equivalent. We therefore have to determine which boson SPT state one get by confining the fermion state. The construction of symmetric gapped surface state (with topological order) outlined in Appendix \ref{b} shows that the Kramers' fermions in the $e_fTm_fT$ topological order couples to the $\mathbb{Z}_2$ gauge field, hence they must be confined with the parton fermion and form charge-$1$ non-Kramers bosons. The deconfined surface state is thus the $eCmC$ topological order, with both $e$ and $m$ being charge-$1$ non-Kramers bosons, which is exactly the surface state of the boson TI.

However, the above result leaves one question unsolved: since the two states $e_fTm_fT$ (topological paramagnet) and $eCmC$ (BTI) are equivalent in fermion systems, why would confinement prefer one state over the other? To answer this question, we need to examine the dynamics of the $\mathbb{Z}_2$ gauge field coupled to the fermions more carefully. We show in Appendix \ref{d} that there are two distinct confinement transitions one can drive the system through: a conventional one resulting from a trivial dynamics of the gauge field (which was implicitly assumed above), and a ``twisted'' one which requires nontrivial dynamics on the gauge field. In our case, the conventional confinement results in the BTI state, while the twisted confinement results in the $e_fTm_fT$ topological paramagnet. Therefore the BTI state seems to be the more natural confined phase, since it only requires a trivial dynamics on the $\mathbb{Z}_2$ gauge field.

\textbf{Acknowledgement}: I thank T. Senthil and A. C. Potter for previous collaborations that led to this work. I am very grateful to T. Senthil and L. Fu for encouragements and helpful discussions. I also thank L. Fu for suggesting the term ``clustons", X. J. Liu for teaching me Efimov state, and E. Tang and T. H. Hsieh for useful comments on the manuscript. This work was supported by NSF DMR-1305741.

\appendix


\section{Hall conductance of fermion SPT states}
\label{0}

Consider a $2D$ system of charged fermions, and assume that there is no fractionalized anyon excitations in the bulk. It is then well known that such a state must have integer hall conductance in unit of $e^2/h$: this can be shown easily by examining the statistics of a $2\pi$ magnetic flux. 
We can then combine the system with some integer quantum hall state to produce a new state with zero hall conductance. Since integer quantum hall states of fermions have equal charge and thermal hall conductance, the combined state will have a new thermal hall conductance
\be
\label{diff2}
\tilde{\kappa}_{xy}=\kappa_{xy}-\(\frac{\pi^2k_B^2T}{3e^2} \)\sigma_{xy},
\ee
where $\kappa_{xy}$ and $\sigma_{xy}$ are the thermal and charge hall conductance of the original state before combining with any integer quantum hall state. Therefore in order to prove Eq.~\eqref{diff}, it is sufficient to prove that the thermal hall conductance must be an integer multiple of eight if the system is non-fractionalized and has zero charge hall conductance.

We now consider the edge state of this system, which in general is a multi-component Luttinger liquid
\be
\mathcal{L}=\frac{1}{4\pi}\left(K_{IJ} \p_x \phi_I \p_t \phi_J +.......\right) +  \frac{1}{2\pi} \epsilon_{\mu\nu} \tau_I \p_\mu \phi_I A_\nu
\ee
described by a symmetric integer $K$-matrix with an integer charge vector $\tau$. A local object will carry odd charge {\em iff} it is a fermion, therefore the parity of the $n$th diagonal element of $K$ must agree with the parity of the $n$th entry of $\tau$. For non-fractionalized bulk, we have $|{\rm det}(K)|=1$.

To make our discussion self-contained, we summarize some key facts known about the edge theory: the operator $e^{il_I\phi_I}$ defined by the integer vector $l$ carries spin $S=\frac{1}{2}l^TK^{-1}l$, and charge $Q=\tau^TK^{-1}l$. It could condense on the edge only if $S=0$. 
When it condenses, another mode (defined by another integer vector $l'$) can stay gapless on the edge only if the operator $e^{il'_I\phi_I}$ commutes with $e^{il_I\phi_I}$, which means $l^TK^{-1}l'=0$. 

The fact that the state has zero charge hall conductance means that 
\be
\sigma_{xy}=\tau^TK^{-1}\tau=0.
\ee
Therefore the operator $e^{i\tau_I\phi_I}$ carries zero spin and charge. We can then introduce a charge-conserving term
\be
\label{chargegap}
\Delta\mathcal{L}=U\cos(\tau_I\phi_I),
\ee
which at sufficiently large $U$ will gap out all the modes that do not commute with it. The remaining gapless modes, denoted as $\tilde{\phi}_{\alpha}=l_{\alpha,I}\phi_I$, must satisfy $\tau^TK^{-1}l=0$ in order to commute with the condensed operator. But this precisely means that the remaining modes are charge-neutral, since $\tau^TK^{-1}l$ gives the charge carried by the operator $e^{il_I\phi_I}$. 

Therefore the remaining edge state can be described using another $\tilde{K}$-matrix after proper field redefinition, with zero charge-vector $\tilde{\tau}=0$ since all modes are charge-neutral. 
Since a local object can be charge-neutral only if it is bosonic, the $\tilde{K}$-matrix must describe a bosonic topological state. In particular, the diagonal elements of $\tilde{K}$ must be even integers. 
It is known\cite{luashvin} that for bosonic states, the minimal non-fractional chiral phase (with a non-zero chiral central charge or thermal hall conductance) is the so-called $E_8$ state, which has $\kappa=8$. The corresponding $\tilde{K}$-matrix of such a state is the Cartan matrix for the exceptional Lie group $E_8$. This proves our assertion under Eq.~\eqref{diff2}, hence also proves Eq.~\eqref{diff}.

As a side note, the above derivation is valid for general abelian topological orders described by a $K$-matrix with $|{\rm det}|\geq1$, with the only modification that the charge gap in Eq.~\eqref{chargegap} should be replaced by $\Delta\mathcal{L}=U\cos(N\tau_I\phi_I)$, where $N$ is some integer that makes the operator local. 
The conclusion is unchanged: if $\sigma_{xy}=0$ and there is no other symmetry than the charge $U(1)$, then the charge modes on the edge can be gapped, and the remaining modes can be described by a charge-neutral bosonic topological order.

\section{The 2D equivalence from edge theories}
\label{a}

We show here the equivalence between the fermion model in Eq.~\eqref{13chern} and the BIQHE state. It is sufficient to show that the boundary between the two states can be fully gapped while preserving charge conservation. The boundary Luttinger liquid
\be
\mathcal{L}=\frac{1}{4\pi}\left(K_{IJ} \p_x \phi_I \p_t \phi_J +.......\right) +  \frac{1}{2\pi} \epsilon_{\mu\nu} \tau_I \p_\mu \phi_I A_\nu
\ee
is described by the $K$-matrix with the charge vector $\tau$:
\be
K=\lp\begin{array}{cc}
K_1 & 0 \\
0 & K_2 \end{array}\rp=\lp\begin{array}{cccc}
          1 & 0 & 0 & 0 \\
          0 & -1 & 0 & 0 \\
          0 & 0 & 0 & 1 \\
          0 & 0 & 1 & 0 \end{array}\rp, 
           \tau=\lp\begin{array}{c}
                                                                         1 \\
                                                                         3 \\
                                                                         2 \\
                                                                         2 \end{array}\rp.
\ee

Now consider two possible mass terms:
\bea
\label{gap}
\Delta\mathcal{L}&=&U_1\cos\Phi_1+U_2\cos\Phi_2 \nonumber \\
&=&U_1\cos\lp\phi_1-\phi_2+\phi_3\rp \nonumber \\
&&+U_2\cos\lp\phi_1+\phi_2-2\phi_4\rp,
\eea
which obviously preserve the $U(1)$ charge conservation. The null vector criteria\cite{haldane} $\Phi_iK^{-1}\Phi_j=0$ is easily satisfied. Therefore Eq.~\eqref{gap} fully gaps out the edge theory while preserving the $U(1)$ symmetry when the couplings $U_{1,2}$ are large.

\section{The 3D equivalence from symmetric surface states}
\label{b}
We show here the equivalence between the fermion model in Eq.~\eqref{13TI} and the Cooper pair boson TI (BTI) state. One of the defining features\cite{avts12,hmodl,metlitski} of the BTI state is that when the surface breaks $U(1)$ but not $\mathcal{T}$, it is gapped without topological order, but the vortex of the surface superconductor has fermion statistics. To access a fully symmetric surface state, one can imagine driving a surface phase transition and condensed double-vortex (which is a boson). It is well known that double-vortex condensates produce $\mathbb{Z}_2$ topological orders\cite{z2long,bfn}, and in our case we get precisely the surface topological orders studied in\cite{avts12,hmodl,metlitski}.

It is easy to see that when $U(1)$ symmetry is broken on the surface, the surface Dirac theory Eq.~\eqref{2dirac} (for general odd $e^*/e=n$) can be fully gapped by introducing the pairing term:
\be
\Delta\mathcal{L}=i\Delta\psi\sigma_y\psi+i\xi\Delta^{n}\Psi\sigma_y\Psi+h.c.,
\ee  
where we wrote the second pairing amplitude as proportional to $\Delta^{n}$ to keep the pairing field $\Delta$ formally charge-$2$, and $\xi$ is a non-universal coupling constant. We then have to show that the vortex in $\Delta$ field has fermion statistics for $n=\pm3 ({\text mod } \hspace{2pt}8)$ and boson statistics for $n=\pm1 ({\text mod } \hspace{2pt}8)$.

Since there are even numbers of Dirac cones in total, the vortex\cite{fkmajorana} does not trap any Majorana zero-mode and is thus abelian. The abelian part of the statistics is then given by the topological spin $e^{i\theta}$, which receives nontrivial contribution from both Dirac cones. It is important to notice here that while the charge-$1$ fermion $\psi$ sees the fundamental vortex as a $\pi$-flux, the charge-$n$ fermion $\Psi$ sees it as a $n\pi$-flux. Therefore the topological spin of the fundamental vortex is given by $e^{i\theta}=e^{i\theta_1}e^{i\theta_n}$, where $e^{i\theta_n}$ is the topological spin of a $n\pi$-flux seen by a paired single Dirac cone. Fortunately this topological spin $e^{i\theta_n}$ has been computed already in Ref. \onlinecite{fSTO1,fSTO2}, and is given by 
\bea
e^{i\theta_n}=1\hspace{5pt} &{\rm if}& \hspace{5pt} n=\pm1 \hspace{2pt} ({\text mod } \hspace{2pt}8), \nn
e^{i\theta_n}=-1\hspace{5pt} &{\rm if}& \hspace{5pt} n=\pm3 \hspace{2pt} ({\text mod } \hspace{2pt}8).
\eea
Therefore, when $n=\pm1 \hspace{2pt} ({\text mod } \hspace{2pt}8)$the total topological spin is $e^{i\theta}=1$ and the fundamental vortex is a boson, and when $n=\pm3 \hspace{2pt} ({\text mod } \hspace{2pt}8)$ the total topological spin is $e^{i\theta}=-1$ and the fundamental vortex is a fermion. When the vortex is boson, it can be condensed and produce a trivial insulator on the surface, and the corresponding bulk state is also trivial. But if the vortex is a fermion, one can no longer condense it to produce a trivial surface insulator. One can instead condense double-vortex to produce an insulator which has intrinsic $\mathbb{Z}_2$ topological order. Such a topological order contains the remnant of the uncondensed vortex $\epsilon$ which is a non-Kramers fermion, and the remnant of the Bogoliubov quasi-particle $e_f$ which is a Kramers fermion. The two particles see each other as $\pi$-flux, and the bound state of the two (denoted as $m_f$) is another fermion which is also Kramers. Therefore the $\mathbb{
Z}_2$ gauge theory contains three distinct fermions, two of which are Kramers. This is exactly the surface topological order of the topological paramagnet $e_fTm_fT$. Binding a physical fermion (charge-$1$, Kramers) to $e_f$ and $m_f$ converts them to charge-$1$ bosons. Thus the topological order can also be viewed as one with two bosons and one fermions, with both bosons carrying charge-$1$. This is exactly the surface topological order of the Cooper pair BTI.

\section{Gapping out cluston helical modes}
\label{c}
Here we show that two copies of helical modes in $1D$, one carrying charge-$1$ and the other one carrying charge-$3$, can be fully gapped without breaking the $U(1)\rtimes\mathcal{T}$ symmetry. Such helical theory arises both in the $\mathbb{Z}_2$ flux core of the fermion system described by Eq.~\eqref{13TI}, and on the edge of a $2D$ state, which is a combination of the free fermion and the cluston $2D$ TI.

The Luttinger liquid
\be
\mathcal{L}=\frac{1}{4\pi}\left(K_{IJ} \p_x \phi_I \p_t \phi_J +.......\right) +  \frac{1}{2\pi} \epsilon_{\mu\nu} \tau_I \p_\mu \phi_I A_\nu,
\ee
is described by the $8\times8$ $K$-matrix: 
\be
K=\lp\begin{array}{cccc}
K_1 & 0 & 0 & 0 \\
0 & K_2 &0 & 0\\
0 & 0 & K_3 & 0 \\
0 & 0 & 0 & K_4 \end{array}\rp=\lp\begin{array}{cccc}
          \sigma_z & 0 & 0 & 0 \\
          0 & \sigma_z & 0 & 0\\
          0 & 0 & \sigma_z & 0\\
          0 & 0 & 0 & \sigma_z\end{array}\rp,          
\ee
charge vector $\tau$:
\be
 \tau=\lp\begin{array}{c}
                                                                         3 \\
                                                                         3 \\
                                                                         1 \\
                                                                         1 \\
                                                                         1 \\
                                                                         1 \\
                                                                         1 \\
                                                                         1\end{array}\rp,
\ee
and time-reversal implementation:
\be
\mathcal{T}:\lp\begin{array}{c}
             \phi_1 \\
             \phi_2 \\
             \phi_3 \\
             \phi_4 \\
             \phi_5 \\
             \phi_6 \\
             \phi_7 \\
             \phi_8
            \end{array}\rp \to \lp\begin{array}{cccc}
-\sigma_x & 0 & 0 & 0 \\
          0 & -\sigma_x & 0 & 0\\
          0 & 0 & -\sigma_x & 0\\
          0 & 0 & 0 & -\sigma_x\end{array} \rp\lp\begin{array}{c}
         \phi_1 \\
             \phi_2 \\
             \phi_3 \\
             \phi_4 \\
             \phi_5 \\
             \phi_6 \\
             \phi_7 \\
             \phi_8
            \end{array}\rp+\pi\lp\begin{array}{c}
0 \\ 1 \\ 0 \\ 1 \\ 0 \\ 1 \\ 0 \\ 1\end{array}\rp.
\ee
The helical modes $K_1=\sigma_z$ comes from the charge-$3$ cluston TI, while $K_{2,3,4}=\sigma_z$ come from the charge-$1$ TI. We choose to work with three charge-$1$ helical modes instead of only one to avoid subtleties from one-band picture.

Now consider the following term:
\bea
\label{gap2}
\Delta\mathcal{L}&=&U\cos\Phi_1+U\cos\Phi_2+U'\cos\Phi_3+U''\cos\Phi_4 \nonumber \\
&=&U\cos\lp-\phi_1+\phi_3+\phi_6+\phi_8\rp \nn
&&+U\cos\lp-\phi_2+\phi_4+\phi_5+\phi_7\rp \nonumber  \\
&&+U'\cos\lp\phi_1-\phi_2+\phi_3-\phi_4\rp \nn
&&+U''\cos\lp\phi_1-\phi_2+\phi_5-\phi_6\rp.
\eea
It is straightforward to check that Eq.~\eqref{gap2} preserves both $U(1)$ and $\mathcal{T}$ symmetry. The null vector criteria\cite{haldane} $\Phi_iK^{-1}\Phi_j=0$ is easily satisfied, so Eq.~\eqref{gap2} fully gaps out the flux core. 
To ensure that the theory does not break $\mathcal{T}$ spontaneously, we should also check the primitivity condition proposed in Ref. \onlinecite{levin}, by checking the mutual primitivity of all the $4\times4$ minors of $\{\Phi_i\}$. This can be done straightforwardly, and indeed the primitivity condition is satisfied.

\section{Conventional and ``twisted'' confinement}
\label{d}

The question can be simplified by considering the combined state of the $e_fTm_fT$ (topological paramagnet) and $eCmC$ (BTI), which has a surface $\mathbb{Z}_2$ topological order with both $e$ and $m$ being electron-like (Kramers fermion carrying charge-$1$) and is therefore dubbed $e_fCTm_fCT$. The equivalence of the two states in fermion systems implies that their combined state is equivalent to a trivial fermion insulator, as can be seen directly from the surface topological order: one can condense the composite of the $e$ particle and the microscopic fermion (or parton in gauged systems), which confines the surface completely without breaking any symmetry.

Our question now becomes: can one get the $e_fCTm_fCT$ state in a boson system, by confining fermions in a trivial insulator coupling to a $\mathbb{Z}_2$ gauge field? Naively the confinement should simply lead to a trivial boson state since the underlying fermion state is trivial. However, we will show below that the nontrivial state can indeed be obtained if the dynamics of the gauge field is sufficiently nontrivial (which, crucially, does not require non-triviality of the underlying fermions state). We will focus our discussion on the specific example, though generalizations to other symmetries/systems are straightforward.

In the presence of time-reversal symmetry $\mathcal{T}$, the $\mathbb{Z}_2$ gauge flux loops can proliferate in different ways, hence giving rise to distinct confined phases. There is always a trivial confinement, namely the flux loops proliferate with a positive-definite amplitude for all configurations, written schematically as
\be
\la W_C\ra\sim 1,
\ee
where $W_C$ is the loop creation operator for configuration $C$. There is however another way of loop proliferation, namely the amplitude acquires a minus sign whenever the loops self-link (one needs to frame the loops into ribbons to make the self-linking well-defined):
\be
\la W_C\ra\sim(-1)^{L_C},
\ee
where $L_{C}$ is the self-linking number of configuration $C$. It was shown in Ref. \onlinecite{xusenthil} that the end point of such loops on the surface become fermions. If the surface was trivial before confinement, after the "twisted" confinement there will be a $\mathbb{Z}_2$ gauge theory emerging on the surface. The excitations of the surface $\mathbb{Z}_2$ gauge theory include the un-condensed flux which is now a fermion, and the deconfined $\mathbb{Z}_2$ gauge charge which only lives on the surface. In our example, the gauge charge is electron-like (fermion carrying $U(1)$ charge and Kramers' degeneracy), and the gauge flux is a non-Kramers fermion carrying no charge (call it $\epsilon$). The surface $\mathbb{Z}_2$ gauge theory therefore has both $e$ and $m$ particles being electron-like. The confined state is thus nothing but the $e_fCTm_fCT$ boson SPT state.





\begin{thebibliography}{99}

 \bibitem{TIs} M. Z. Hasan and C. L. Kane, Rev. Mod. Phys. 82, 3045 (2010). X.-L. Qi and S.-C. Zhang, Rev. Mod. Phys. 83, 1057 (2011). M. Z. Hasan and J. E. Moore, Annu. Rev. Condens. Matter Phys. 2, 55 (2011).
 
  \bibitem{tenfold} Alexei Kitaev, http://arxiv.org/abs/0901.2686. Shinsei Ryu, Andreas P Schnyder, Akira Furusaki and Andreas W W Ludwig, New J. Phys. 12 065010 (2010), X.G. Wen, Phys. Rev. B 85, 085103 (2012).
  
\bibitem{review} T. Senthil, arXiv:1405.4015. Ari M. Turner and Ashvin Vishwanath, arXiv:1301.0330

  

\bibitem{Wenbook} X. G. Wen, Quantum Field Theory Of Many-body Systems: From The
  Origin Of Sound To An Origin Of Light And Electrons. Oxford University Press (2004).



\bibitem{chencoho2011} Xie Chen, Zheng-Cheng Gu, Zheng-Xin Liu, Xiao-Gang Wen,
Science 338, 1604 (2012); Phys. Rev. B 87, 155114 (2013).

\bibitem{note0} Note: by ``free fermion state'', what we really mean is an interacting fermion state that is adiabatically connected to a state realized by a free fermion model.


\bibitem{trivialhaldane}F. Anfuso and A. Rosch, Phys. Rev. B 75, 144420 (2007).

\bibitem{3dfSPT} C. Wang, A.C. Potter, and T. Senthil, Science 343, 629 (2014).

\bibitem{TScSTO}Lukasz Fidkowski, Xie Chen, and Ashvin Vishwanath, arXiv:1305.5851. 

\bibitem{luashvin} Yuan-Ming Lu and Ashvin Vishwanath, Phys. Rev. B 86, 125119 (2012).

\bibitem{1dfSPT} L. Fidkowski and A. Kitaev, Phys. Rev. B 81, 134509 (2010); L. Fidkowski and A. Kitaev, Phys. Rev. B 83, 075103 (2011); Evelyn Tang, Xiao-Gang Wen, Phys. Rev. Lett. 109, 096403 (2012).

\bibitem{3dfSPT2} C. Wang and T. Senthil, Phys. Rev. B 89, 195124 (2014).

\bibitem{efimov} T. Kraemer, M. Mark, P. Waldburger, J. G. Danzl, C. Chin, B. Engeser, A. D. Lange, K. Pilch, A. Jaakkola, H.-C. Nägerl and R. Grimm, Nature 440, 315 (2006).




\bibitem{tsml}T. Senthil, Michael Levin,   Phys. Rev. Lett. 110, 046801 (2013).

\bibitem{liuwen}Zheng-Xin Liu and Xiao-Gang Wen, Phys. Rev. Lett. 110, 067205 (2013).



 \bibitem{avts12}A. Vishwanath, T. Senthil,  Phys. Rev. X 3, 011016 (2013). 


 \bibitem{hmodl} C. Wang and T. Senthil, Phys. Rev. B 87, 235122 (2013).


\bibitem{metlitski}Max A. Metlitski, C. L. Kane and Matthew P. A. Fisher, http://arxiv.org/abs/1302.6535.
 

\bibitem{kitaev}A. Kitaev, Ann. Phys. 321, 2-111 (2006).




 
 













\bibitem{FKM} L. Fu, C. L. Kane, and E. J. Mele, Phys. Rev. Lett. {\bf 98}, 106803 (2007).

\bibitem{qi}Xiao-Liang Qi, Taylor L. Hughes, and Shou-Cheng Zhang, Phys. Rev. B 78, 195424 (2008).


\bibitem{paranomaly} A. N. Redlich, Phys. Rev. Lett. 52, 18 (1984); Phys. Rev. D 29, 2366 (1984).

\bibitem{note} Note: more precisely, we have $c-k=4({\rm mod}8)$ where $k$ is the level of the Chern-Simons term of the $U(1)$ gauge field.

\bibitem{graanomaly} Shinsei Ryu, Joel E. Moore and Andreas W. W. Ludwig, Phys. Rev. B 85, 045104 (2012).

\bibitem{2dparton}P. Ye and X.-G. Wen, Phys. Rev. B 87, 195128 (2013);
Y.-M. Lu and D.-H. Lee, arXiv:1210.0909 (2012); T. Grover and A. Vishwanath, Phys. Rev. B 87, 045129 (2013); J. Oon, G. Y. Cho, and C. Xu, arXiv:1212.1726 (2012).

\bibitem{3dparton} P. Ye and X.-G. Wen, arXiv:1303.3572 (2013); Max A. Metlitski, C. L. Kane and Matthew P. A. Fisher, unpublished.


\bibitem{z2defect}Yi Zhang, Ying Ran and Ashvin Vishwanath, Phys. Rev. B 79, 245331 (2009).

\bibitem{kanemele}C.L. Kane and E.J. Mele, Phys. Rev. Lett. 95, 146802 (2005).




\bibitem{haldane} F. D. M. Haldane, Phys. Rev. Lett. 74, 2090 (1995).

\bibitem{levin} Michael Levin and Ady Stern, Phys. Rev. B 86, 115131 (2012).


  \bibitem{fSTO1}Chong Wang, Andrew C. Potter, and T. Senthil,  Phys. Rev. B 88, 115137 (2013). 


\bibitem{fSTO2}Max A. Metlitski,   C. L. Kane, and Matthew P. A. Fisher, arXiv:1306.3286.





 \bibitem{z2long}T. Senthil and Matthew P. A. Fisher, Phys. Rev. B 62, 7850 (2000).





\bibitem{bfn}Leon Balents, Matthew P. A. Fisher, and Chetan Nayak, 
Phys. Rev. B 60, 1654 (1999). 

\bibitem{fkmajorana}Liang Fu and C. L. Kane, Phys. Rev. Lett. 100, 096407 (2008).

\bibitem{z2xu}Zhen Bi, Alex Rasmussen, Yizhuang You, Meng Cheng and Cenke Xu, arXiv:1404.6256

\bibitem{xusenthil}Cenke Xu and T. Senthil, Phys. Rev. B 87, 174412 (2013).

\bibitem{avdomain}Xie Chen, Yuan-Ming Lu, Ashvin Vishwanath,  arXiv:1303.4301. 


\end{thebibliography}
\end{document}